# Deuterium Balmer/Stark spectroscopy and impurity profiles: first results from mirror-link divertor spectroscopy system on the JET ITER-like wall

A.G. Meigs[a*], S. Brezinsek[b], M. Clever[b], A. Huber[b], S. Marsen[c], C. Nicholas[d], M.Stamp[a], K-D Zastrow[a], *and JET EFDA Contributors*[#]

JET-EFDA, Culham Science Centre, Abingdon, OX14 3DB, UK
[a]*EURATOM/CCFE Fusion Association, Culham Science Centre, Abingdon, OX14 3DB, UK*
[b]Institute of Energy and Climate Research –Plasma Physics, Forschungszentrum Jülich
[c]*Max-Planck-Institut for Plasma Physics, EURATOM Association, Greifswald, Germany*
[d]Dept. of Physics, University of Strathclyde, Glasgow G4 0NG

For the ITER-like wall, the JET mirror link divertor spectroscopy system was redesigned to fully cover the tungsten horizontal strike plate with faster time resolution and improved near-UV performance. Since the ITER-like wall project involves a change in JET from a carbon dominated machine to a beryllium and tungsten dominated machine with residual carbon, the aim of the system is to provide the recycling flux, equivalent, to the impinging deuterium ion flux, the impurity fluxes (C, Be, O) and tungsten sputtering fluxes and hence give information on the tungsten divertor source. In order to do this self-consistently, the system also needs to provide plasma characterization through the deuterium Balmer spectra measurements of electron density and temperature during high density. L-Mode results at the density limit from Stark broadening/line ratio analysis will be presented and compared to Langmuir probe profiles and 2D-tomography of low-n Balmer emission [1]. Comparison with other diagnostics will be vital for modelling attempts with the EDGE2D-EIRENE code[2] as the best possible data sets need to be provided to study detachment behaviour.

*This work, supported by the European Communities under the contract of Association between EURATOM/CCFE was carried out within the framework of the European Fusion Development Agreement. The views and opinions expressed herein do not necessarily reflect those of the European Commission. This work was also part-funded by the RCUK Energy Programme under grant EP/I501045..*
#See the Appendix of F. Romanelli et al., Proceedings of the 23rd IAEA Fusion Energy Conference 2010, Daejeon, Korea
*JNM keywords: B0100 Beryllium, C 0100 Carbon, D0500 – Divertor materials, F0400 –First wall materials, P0500 Plasma-material interaction*
*PSI-20 keywords: carbon, beryllium, tungsten, JET*
 *Corresponding author address: JET-EFDA, Culham Science Centre, Abingdon, OX14 3DB, UK*
*Presenting author: Andrew Meigs*
*Presenting author e-mail: Andrew.Meigs@ccfe.ac.uk*

**Introduction**

Spatially resolved spectroscopy in the near-UV to near-IR of the outer divertor of JET is extremely important to many different aspects of characterizing and modelling the behaviour of the main ions and impurity ions in the divertor. In addition, the mirror-link diagnostic provides invaluable cross-calibration of the tomographic reconstructions of medium-n Balmer lines from the tangentially viewing filter cameras without which line ratio of D-gamma/D-beta could not be accurately determined.[1]. Being able to measure the influx of tungsten from neutral lines in the blue while simultaneously monitoring the Balmer series, beryllium and carbon lines as well as other impurities means one set of diagnostics can provide much of the input on impurity behaviour in the outer divertor.

**The diagnostic system**

The diagnostic is a mirror-link system with one long focal length fused silica lens as the imaging optic and fused-silica windows for the vacuum port and biological shield entries. The remaining optical components are mirrors or fused silica dichroic filters. This means that the diagnostic can transmit light from 200nm up to around 2.5microns from the plasma to the spectrometers. The near-UV access allows measurement of W I transitions [3] which are not hampered by any thermal continuum radiation, thus, measurement of sputtering yields at high power loads and energies of the outer target plate. This gives an unprecedented wavelength range for a tritium compatible spectroscopic system. The design and upgraded capabilities of the diagnostic were presented at HTPD 2010[4]. The diagnostic is in fact three spectrometers each optimized for different wavelength bands but sharing the same line of sight. Typically, the near-UV and visible system can acquire 21 LOS with 40msec exposure time and 18mm spatial resolution. The near-IR system using an old camera acquires 14 LOS at 125msec pre frame and ~25mm per LOS. Figure 1 shows the lines of sight as they exit the torus and move through the labyrinth of mirrors the collection optics/spectrometers in the torus hall of JET. The optical path is around 35-meters from divertor floor to CCD detector. In Figure 2, two types of intensity plots are shown (axis on the right) with the divertor geometry overlaid (left axis). The black line is the traditional 2-dimensional plot of the intensity of W I 400.8nm is shown in black; the radial coordinate is actually the radius at which the lines of sight would strike the plane of the load bearing plate. The second type shown as color-coded bands of intensity allows one to stress that the diagnostic is sampling emission along the near-vertical line of sights. This view in the author's opinion overlaid with the divertor geometry really gives insight into understanding the emission profiles.

**Balmer/Stark Spectroscopy in the divertor**

Besides impurity emission and influx the diagnostic is optimized to the study of the Stark broadened Balmer and Paschen series in high density discharges. Being able to measure impurity influx on one instrument while simultaneously measuring the Stark broadened Balmer or Paschen series limit spectra could allow self-consistent application of Balmer spectra derived electron density and temperature to determine impurity influx in the divertor. During first operation of the ITER-Like wall at JET [5, 6] several experiments investigating the density limit behaviour of the "new" machine were performed [1]. Figure 3 shows a sample high resolution Balmer series limit spectra from an L-mode density limit discharge (81469, $B_T$=2.5T, $I_p$=2.5MA, $n_e$*dl=18.1x$10^{19}$m$^{-2}$, $T_e$~1.8keV, $P_{rad}$=2.2MW, $P_{NBI}$=1.1MW) before high density (t=8.06s) and during high density (t=13.02s ). This discharge is an outer horizontal strike point plasma in L-mode with low additional heating. Figure 4 show time traces of the standard plasma parameters. The spectra is modelled and fit using the ADAS/ffs framework with application of a simple model for the Balmer line shapes as Voigt profile [7]. Only the 10-2 to 13-2 transitions are included in the fit as a good model of the continuum has not been implemented yet. The Gaussian component of the Voigt is fixed by determining the width of the impurity lines just at the start of observation of the Stark broadened spectra. The widths of the Lorentzian component of the Voigts are coupled to the Griem density formula [8]. The positions of the Balmer lines are coupled to the position of the D10-2 line through application of the ratio of the Rydberg formula. All impurity lines (several O III) lines appear early in the spectra but are ignored in the final fit to no noticeable detriment. The resulting electron density, intensity of the D10-2 line and the line ratio of the 12-2 to the 10-2 are shown as contour plots in figures 5-7. Both the electron density and the intensity of the Balmer lines show similar behaviour and both show the MARFE transition. Cross-section plots for selected time slices are shown in figure 8 which includes the surface Langmuir probe electron density for comparison. The line ratios are rather flat (see figure 7) and first attempts of coupling the line ratios, the electron density and ADAS photon emission coefficient ratios to determine the electron temperature failed. Previous work using the Paschen continuum step showed temperatures in the carbon machine of 1-4eV[9]. Work is ongoing to reproduce the temperature results using both line ratio's and the continuum steps.

Next the Balmer spectral analysis will be applied to a similar discharge but with vertical targets for the strike points and slightly higher auxiliary heating: 81548 ($B_T$=2.4T, $I_p$=1.7MA, $n_e$*dl=13.0x$10^{19}$m$^{-2}$, $T_e$~1.5keV, $P_{rad}$=MW, $P_{NBI}$=2.3MW). The same model was used in fitting this pulse with some initial parameter changes. Figure 7 shows the basic plasma parameters for the pulse. Figure 8 shows the resulting derived electron density. From this one sees that as the MARFE forms for this vertical strike point discharge the peak of the density moves outboard away from the X-point. Figure 9 shows the D10-2 intensity contour. Inversion of wide-angle D-alpha camera data from the divertor yield volume emission centered on the strike points until the MARFE forms then it appears that the emission moves inside the SOL [1] which could be what mirror-link diagnostic sees as a outboard shift since it views vertically through the SOL from above.

**Summary**

The upgraded mirror-link diagnostic has been providing key spectroscopic observations since JET has restarted with the ITER-like wall. The success of the upgrade can be seen not only in the limited results shown here but by the many other papers/poster at this conference using the diagnostics data. The most notable result as far as the upgrade of the diagnostic is concerned is that for the first time the formation of the density limit MARFE can be see from strike point to x-point using Balmer spectroscopy. Preliminary plans to utilize the light lost at each of the current systems Newtonian telescopes are going forward with the aim of implementing a set of fast acquisition detectors using either filters or compact spectrometers. Analysis is ongoing for the data taken so far. The Balmer analysis will be refined using the simple models of Voigts and a better baseline/continuum and hopefully, re-implementation of the Marseille PPP Balmer line shape code into the model will be undertaken [10].

**Acknowledgements**

This work, supported by the European Communities under the contract of Association between EURATOM and CCFE, was carried out within the framework of the European Fusion Development Agreement. The views and opinions expressed herein do not necessarily reflect those of the European Commission. This work was also part-funded by the RCUK Energy Programme under grant EP/I501045

**Figure Captions**

Figure 1. Engineering design drawing showing the torus hall optical mounts and components. The optical distance from divertor floor (not shown) and ccd detector (not shown) is ~35-meters.

Figure 2. Intensity profile of W I 400.8nm line overlaid on the divertor geometry. Two representations are present: a traditional 2D plot of intensity versus major radius (in black) and a filled polygon representation (colored bands). In the polygon representation, the lines of sight are delineated as color-coded bands surrounded by a goldenrod border. Also, Note the diagnostic looks through from the top of the machine; this image is zoomed just into the divertor. This W I profile is from a frame spanning an ELM. The pulse is 82668.

Figure 3. Spectra from pulse 81469 prior to high density (t=8.06s) and during high density (t=13.02s).

Figure 4. Primary plasma parameters for pulse 81469.

Figure 5. Pulse 81469: Electron density derived from the Lorentzian width of the Balmer 10-2 to 13-2 lines along with magnetics outer strike point position in black and the x-point position in purple. High density throughout the density ramp is located at the strike point until the MARFE transition which lasts from ~12s to 13s. A clear stable MARFE is seen from ~13s to ~15.3s.

Figure 6. Pulse 81469: Intensity of the D10-2 line shows similar behaviour as the electron density. However, after MARFE formation the intensity continues to build at the X-point.

Figure 7. Pulse 81469: The intensity ratio of D11-2 to D10-2 does not show much structure.

Figure 8. Pulse 81469: Two-dimensional plots of Balmer/Stark derived electron density versus Langmuir probe electron density. Early in the discharge (t=8.46s) through the density ramp (t=11.74s), the MARFE transition (t=12.98s) and into the MARFE (t=13.98s)

Figure 9. Primary plasma parameters for pulse 81548.

Figure 10. Pulse 81548: Electron density derived from the Lorentzian width of the Balmer 10-2 to 13-2 lines along with magnetics lower x-point position in purple. Around 22s there appears to be a clear separation from the vertical wall (R~2.85m) as the MARFE forms.

Figure 11. Pulse 81548: Intensity of the D10-2 line shows similar behaviour as the electron density.

**Figures**

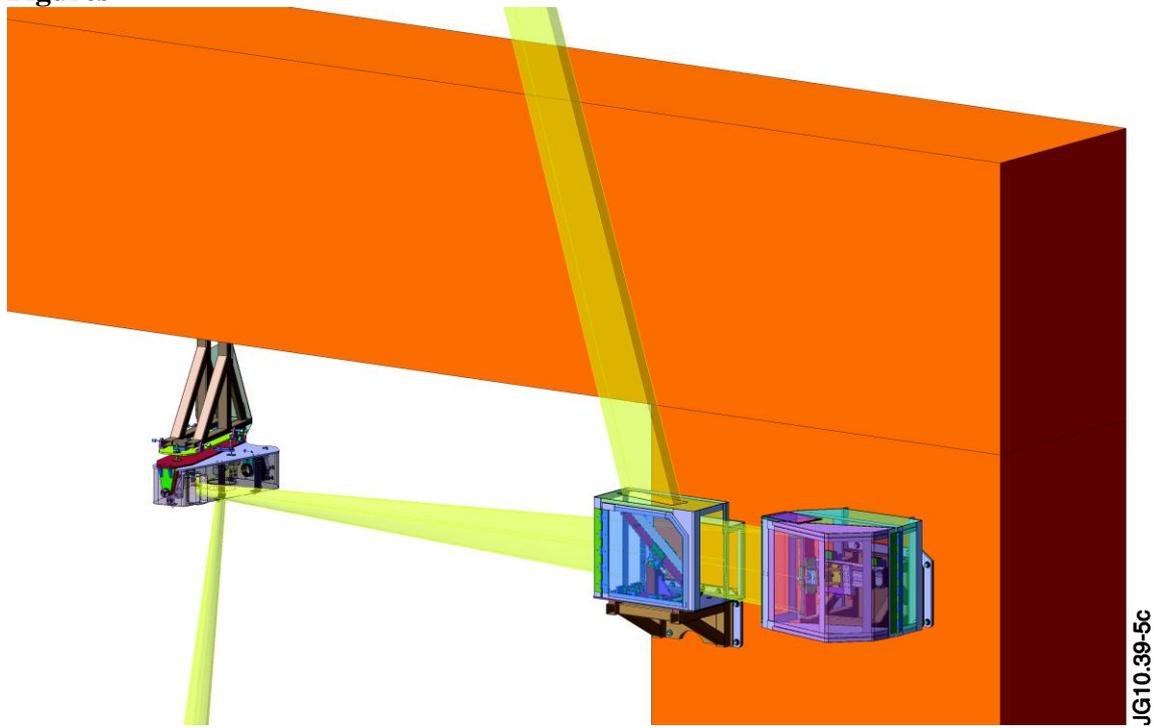

Figure 1. Andrew Meigs, P1-89 PSI2012

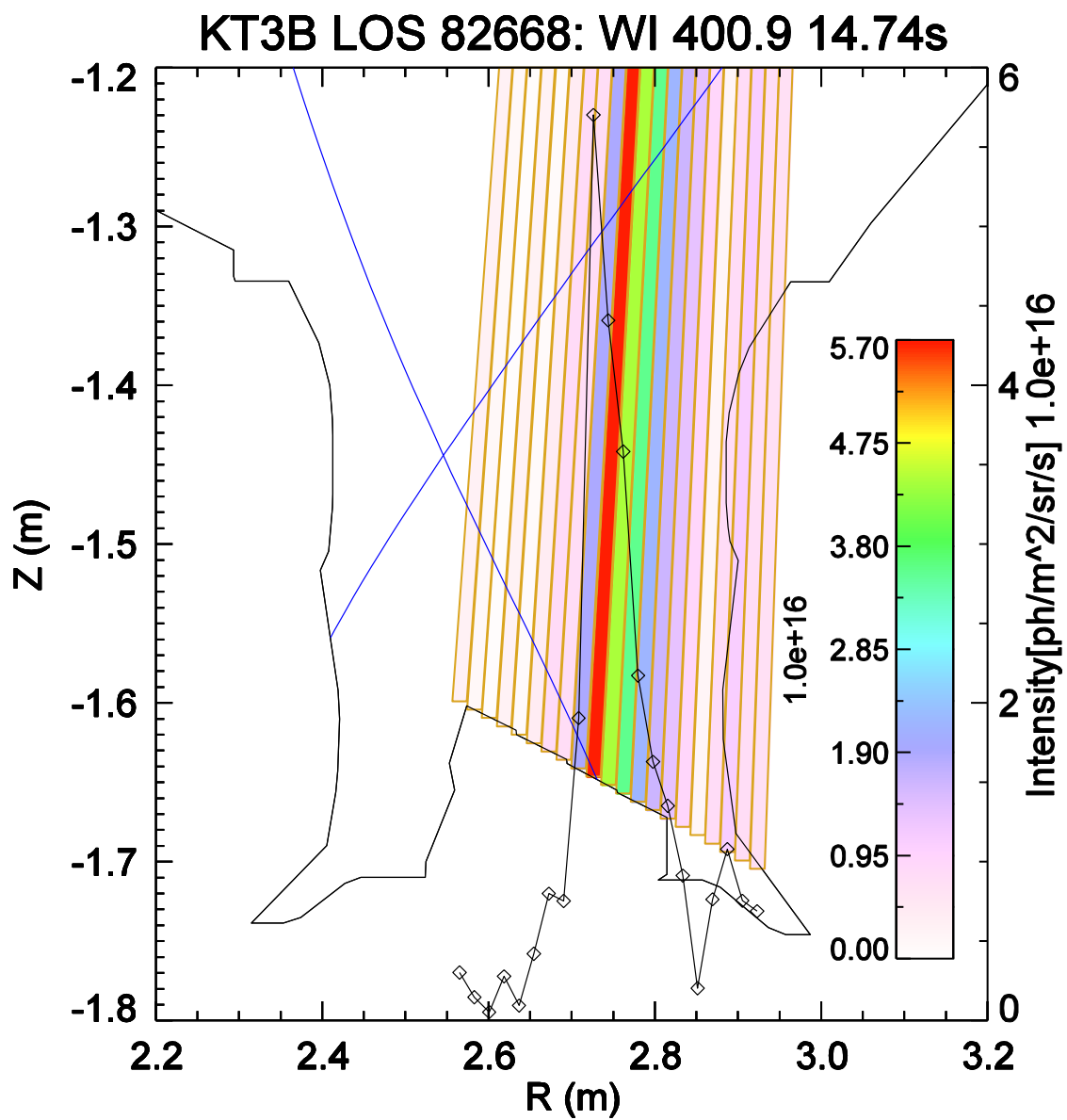

Figure 2. Andrew Meigs, P1-89 PSI2012

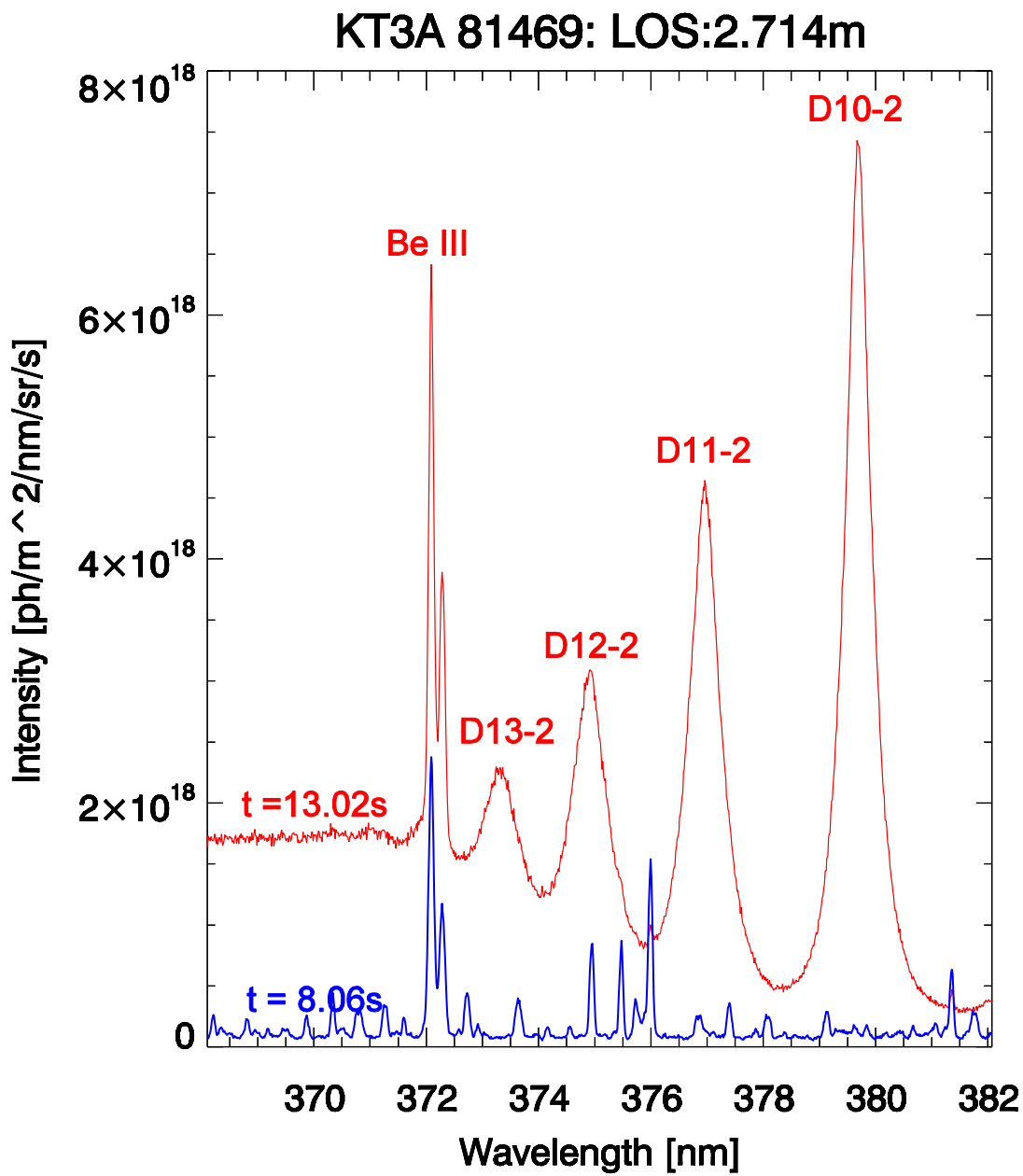

Figure 3. Andrew Meigs, P1-89 PSI2012

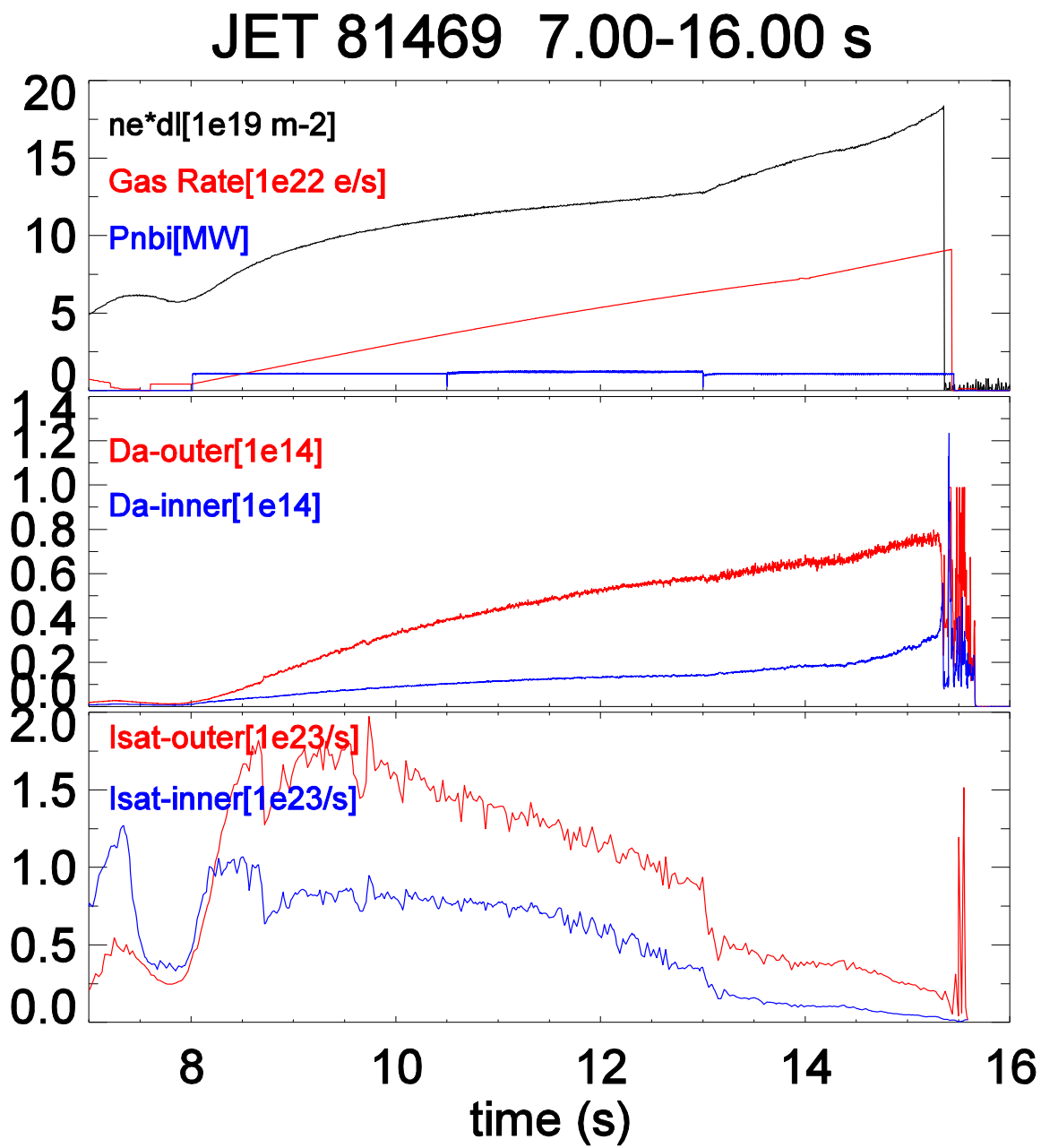

Figure 4. Andrew Meigs, P1-89 PSI2012.

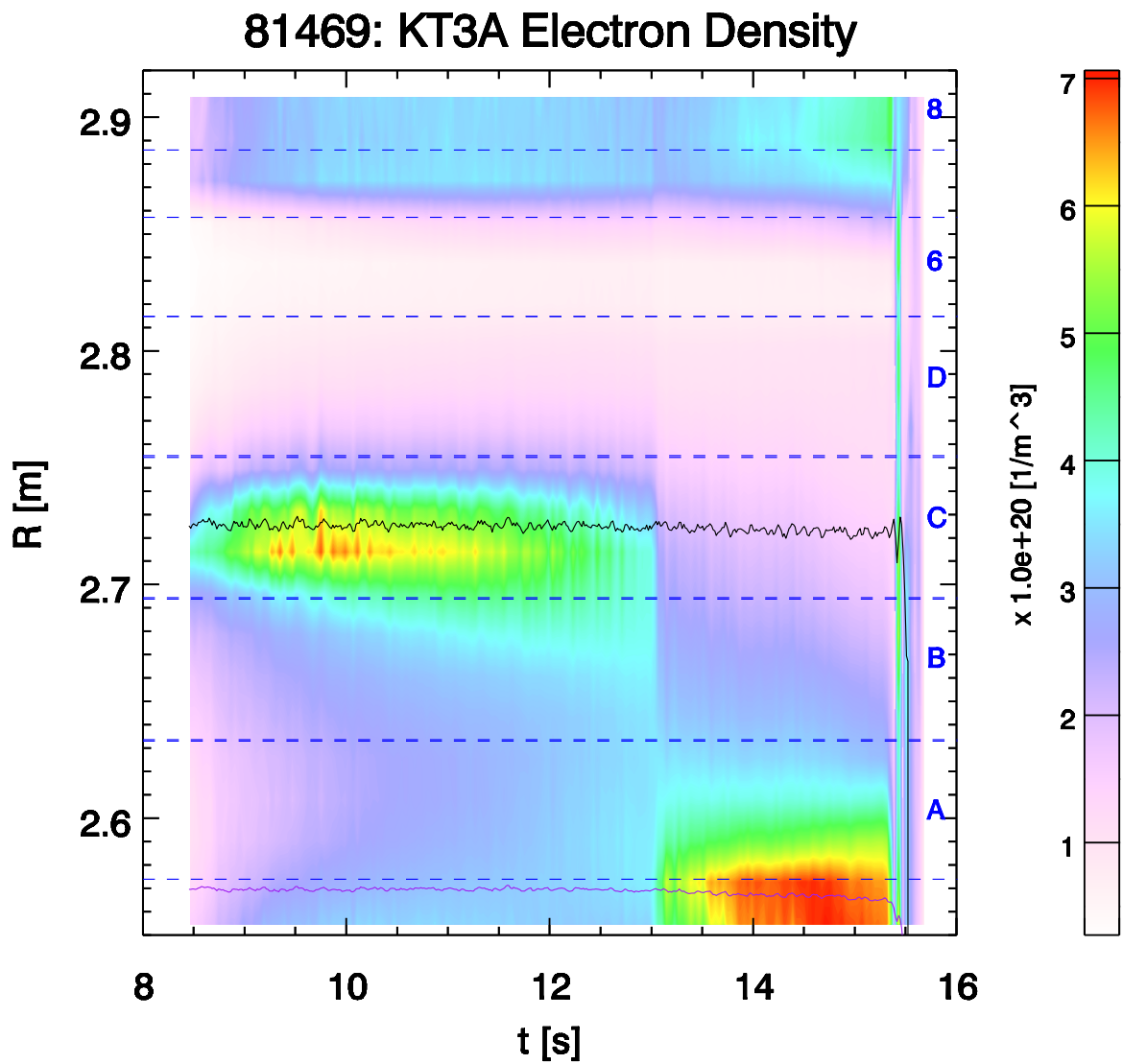

Figure 5. Andrew Meigs, P1-89 PSI2012.

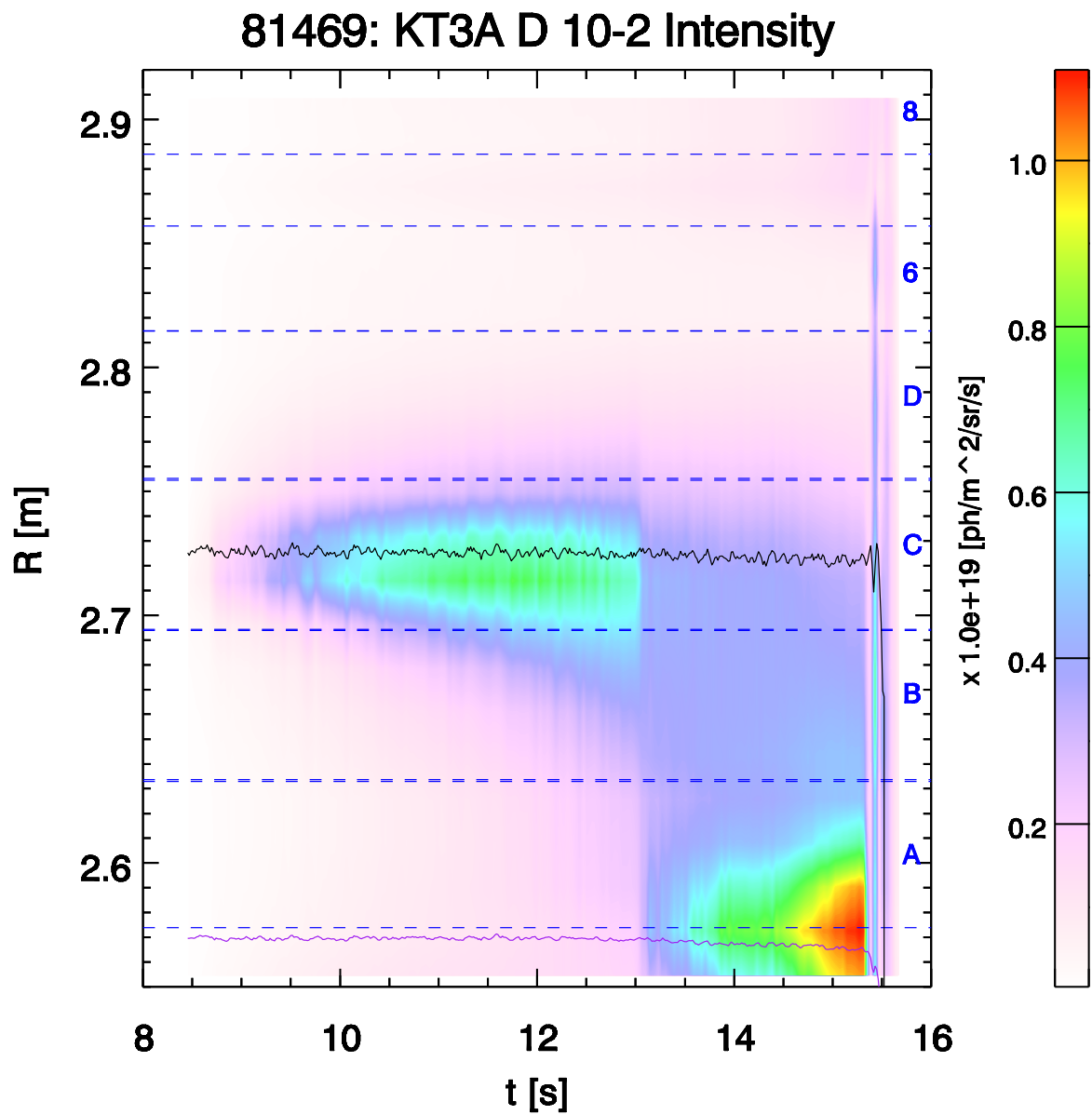

81469: KT3A D 10-2 Intensity

Figure 6. Andrew Meigs, P1-89 PSI2012.

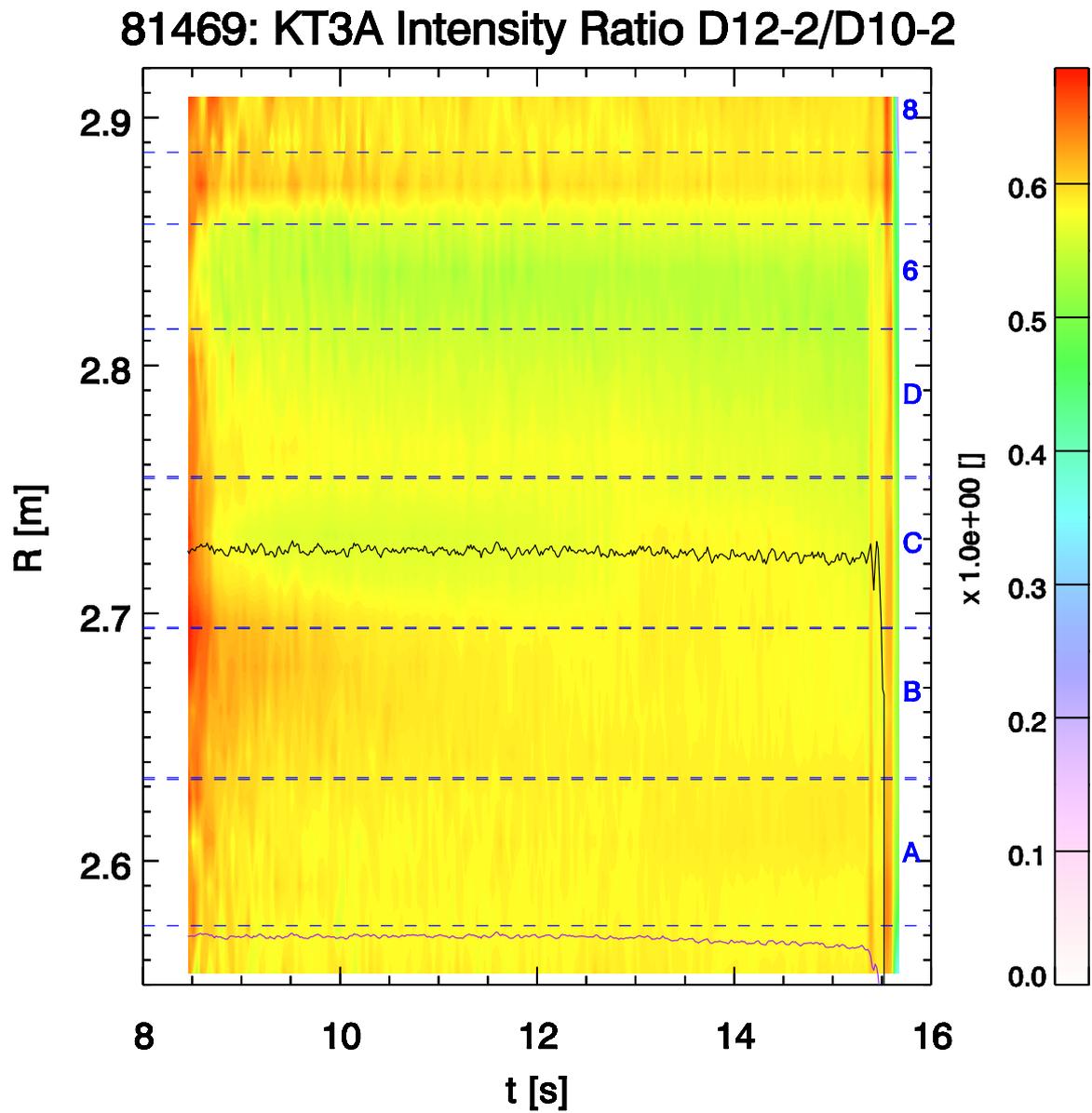

Figure 7. Andrew Meigs, P1-89 PSI2012..

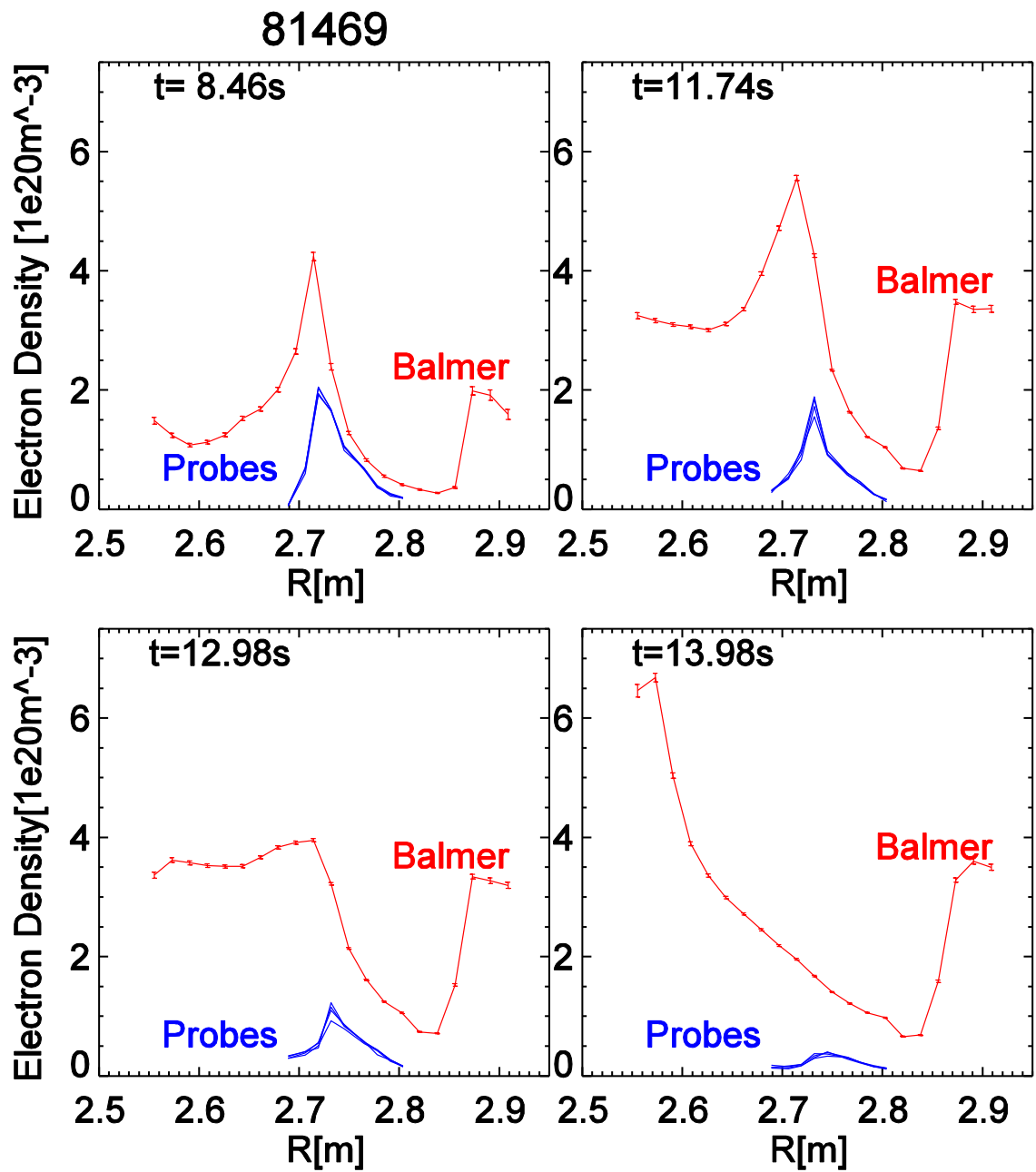

Figure 8. Andrew Meigs, P1-89 PSI2012

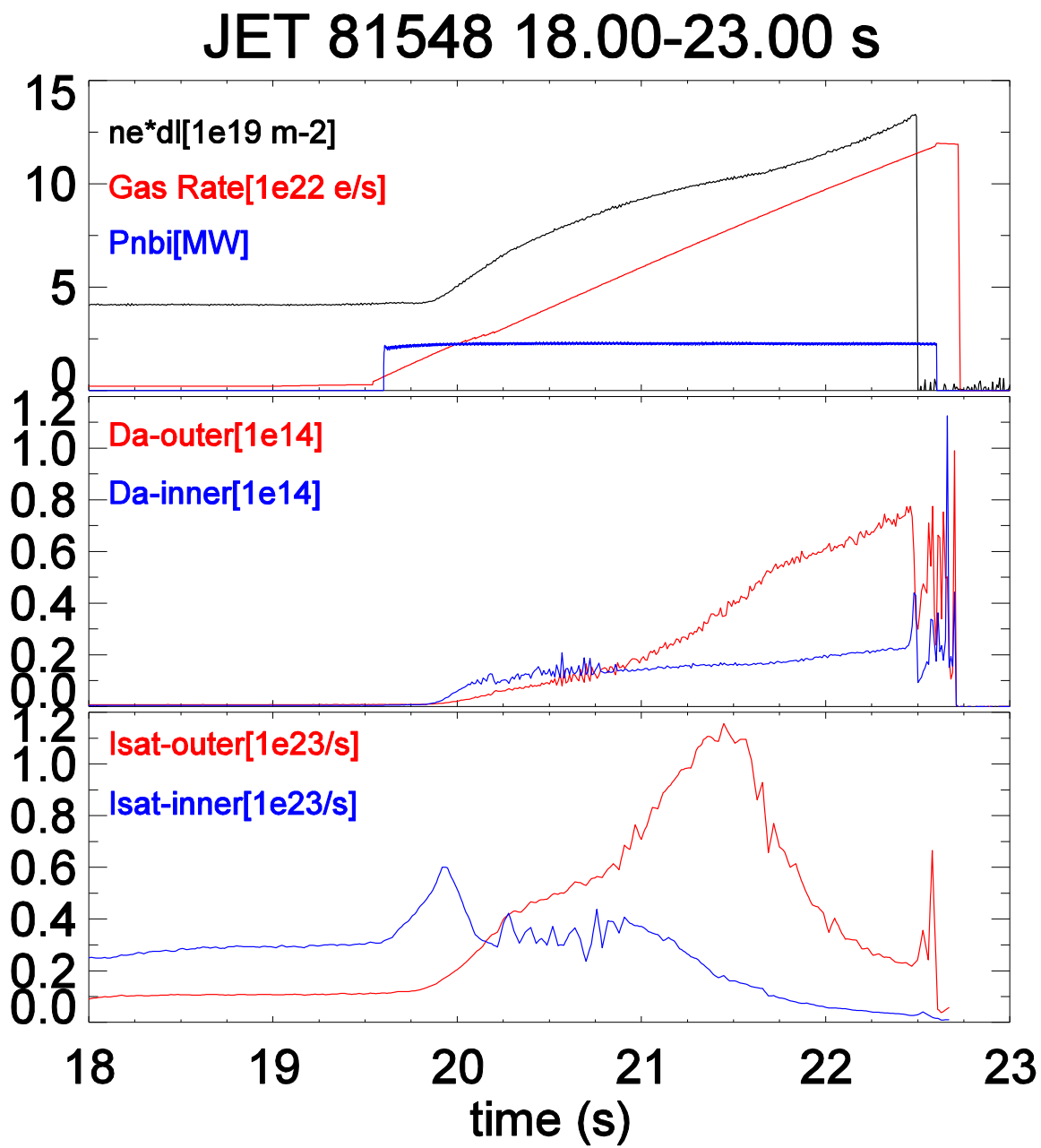

Figure 9. Andrew Meigs, P1-89 PSI2012

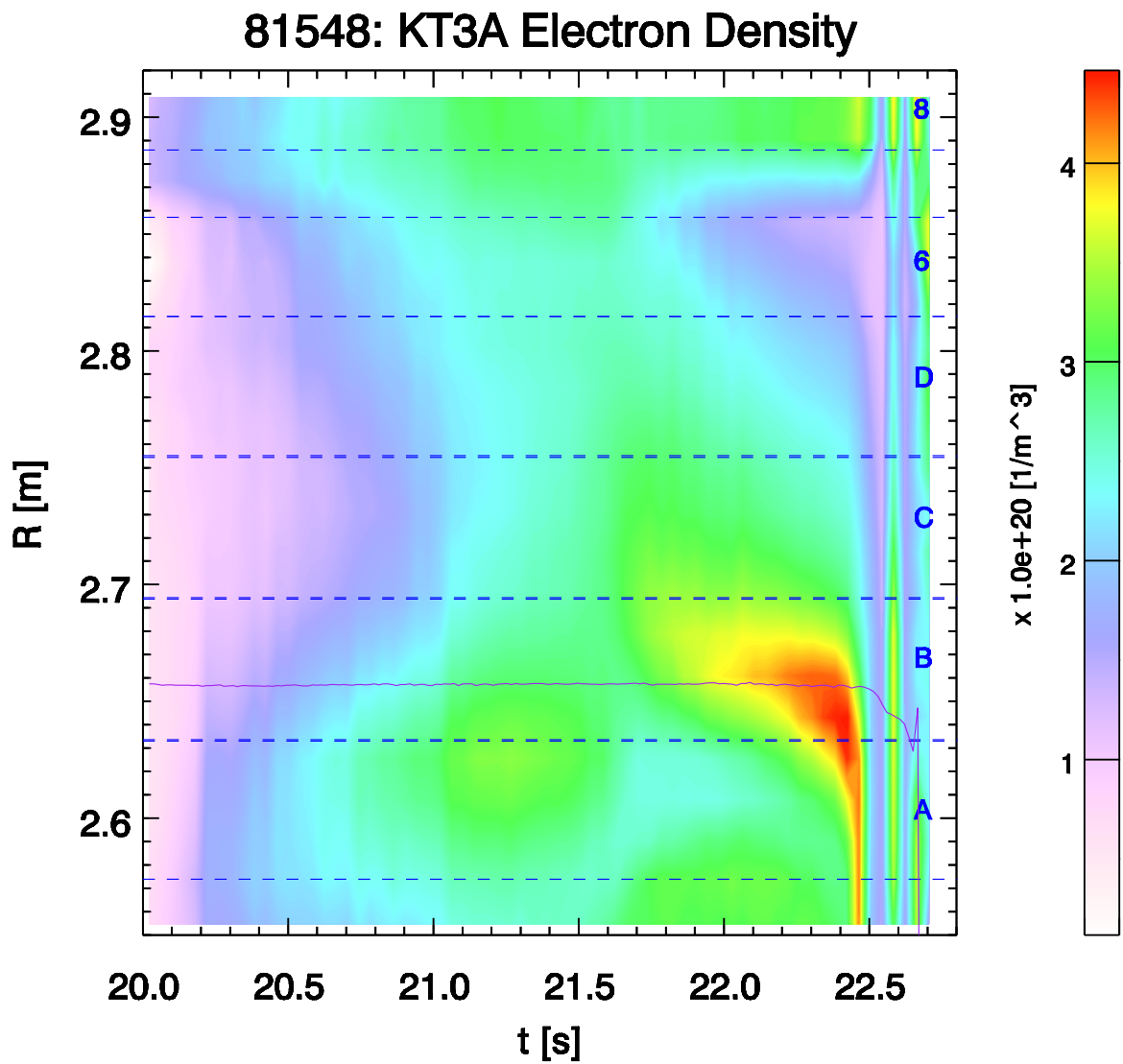

Figure 10. Andrew Meigs, P1-89 PSI2012

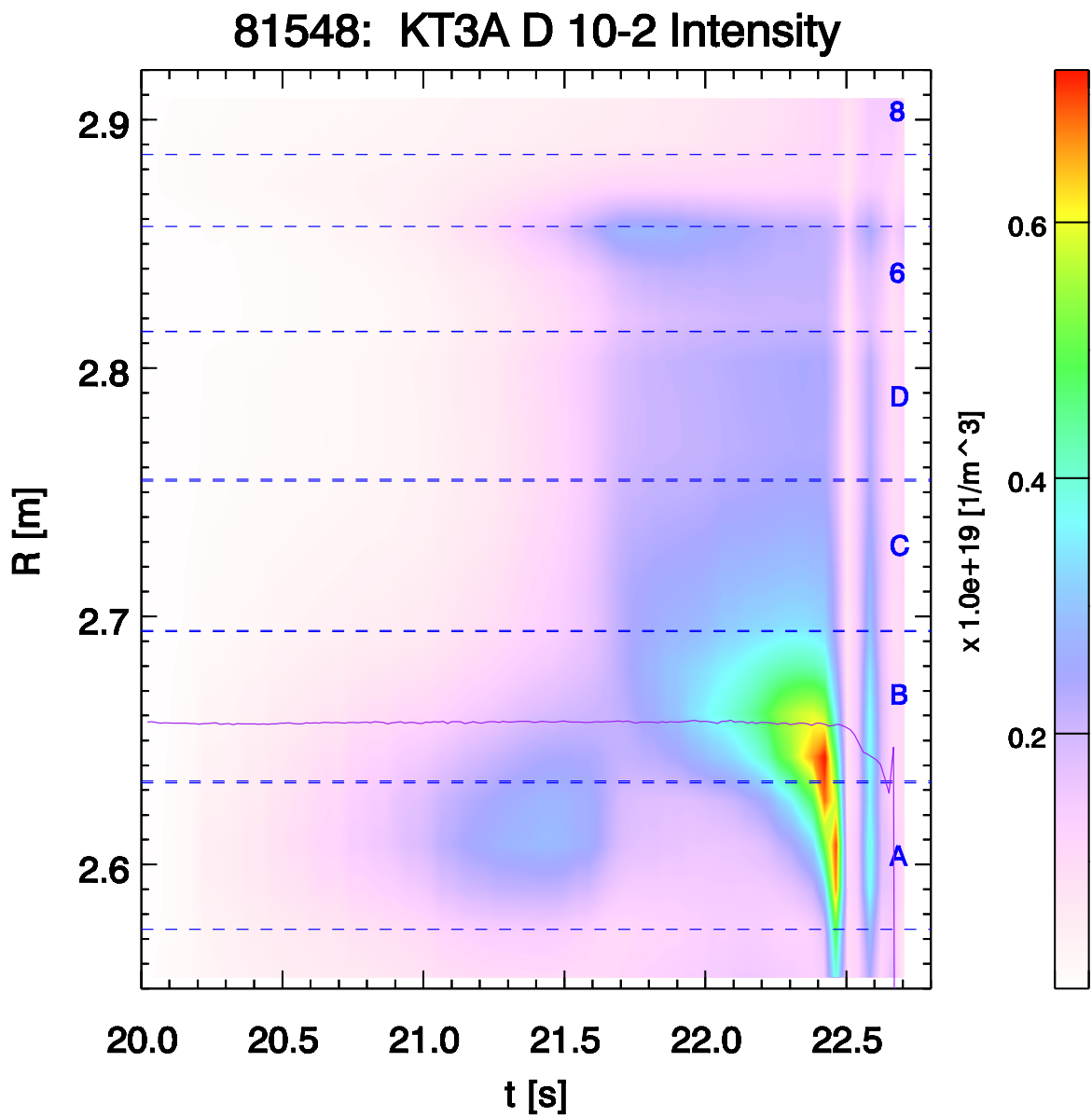

Figure 11. Andrew Meigs, P1-89 PSI2012